\documentclass[useAMS,usenatbib,letterpaper,fleqn]{mn2e}
\usepackage{amsmath}
\usepackage{amsfonts}
\usepackage{amssymb}
\usepackage{graphicx}
\usepackage{threeparttable}
\usepackage{booktabs}
\usepackage{xcolor}
\usepackage{aas_macros}
\usepackage{mathrsfs}
\usepackage[top=1.75cm,bottom=1.75cm,left=2cm,right=2cm]{geometry}
\usepackage[colorlinks=true,linkcolor=blue,citecolor=blue,urlcolor=blue,filecolor=blue]{hyperref}
\usepackage{caption}

\defcitealias{Busha03}{B03}
\defcitealias{Dunner06}{D06}
\defcitealias{Pearson13}{P13}
\defcitealias{Pearson14}{PB$^2$}

\title[The extent of bound structure]{The extent of gravitationally bound structure in a $\Lambda$CDM universe}
\author[D. W. Pearson]{David W. Pearson\thanks{E-mail: \href{mailto:david.pearson@umit.maine.edu}{david.pearson@umit.maine.edu} \newline \href{mailto:dpearson1983@gmail.com}{dpearson1983@gmail.com}} \\
Department of Physics and Astronomy, University of Maine, 120 Bennett Hall, Orono, ME 04469, USA}
\begin{document}

\date{Accepted 2015 March 7. Received 2015 January 28; in original form 2014 August 19}

\pagerange{\pageref{firstpage}--\pageref{lastpage}} \pubyear{2014}

\maketitle

\label{firstpage}

\begin{abstract}
A new analytical model for constraining the extent of gravitationally bound structure in the Universe is presented. This model is based on a simple modification of the spherical collapse model (SCM), and its performance in predicting the limits of bound structure in $N$-body simulations is compared to that of two previous models with the aid of new software named \textsc{coldg}a\textsc{s} -- compute unified device architecture (CUDA) object location determination in \textsc{gadget2} snapshots -- which was developed by the author. All of these models can be distilled down to a single unique parameter $\xi$, here named the critical parameter, which was found to have values of 3 and 1.18 from the previous studies, and a value of 1.89 from the modified SCM. While still on the conservative side, this new model tends to better identify what structure is gravitationally bound in simulations. All three analytical models are applied to the Corona Borealis supercluster, with the modified SCM and \mbox{$\xi = 1.18$} model making predictions that are in agreement with recent work showing that A2056, A2061, A2065, A2067, and A2089 comprise a gravitationally bound supercluster. As an additional test, the modified SCM is used to estimate the mass within the turn around radius of the Virgo cluster, providing results in good agreement with studies relating the virial mass of clusters to the total mass within turn around.
\end{abstract}


\section{Introduction}
With the release of the \textit{Planck} survey data \citep{Ade2014}, we have entered an unprecedented era of precision cosmology. The values of the cosmological parameters from that survey are in excellent agreement with the standard $\Lambda\mathrm{CDM}$ cosmology, meaning that it is all but certain that we live in a universe that has entered a phase of accelerated expansion. This will cause gravitationally bound structures, such as groups and clusters of galaxies as well as dense superclusters, to one day become `island universes' \citep{Nagamine2003,Araya-Melo2009}, with other structures eventually expanding beyond the observable horizon.

Locating the largest gravitationally bound structures has implications beyond simply defining objects destined to become `island universes'. Accurately determining the number of extreme density peaks could place constraints on the nature of the primordial fluctuation field \citep{Sheth2011,Park2012}. So far, there are at least three extreme density peaks known within \mbox{$z \sim 0.2$}, the Shapley supercluster (SSC; \citeauthor{Bardelli1993} \citeyear{Bardelli1993}; \citeauthor{Proust06} \citeyear{Proust06}), the Sloan Great Wall \citep{Luparello2011}, and the Corona Borealis supercluster (CSC; \citeauthor{Small98} \citeyear{Small98}; \citeauthor{Luparello2011} \citeyear{Luparello2011}; \citeauthor{Batiste13} \citeyear{Batiste13}) all with estimated masses of \mbox{$\sim 10^{16} \, h^{-1} \, \mathrm{M}_{\sun}$}. Recently \citeauthor{Pearson14} (\citeyear{Pearson14}, hereafter \citetalias{Pearson14}) provided the most conclusive evidence to date that the CSC contains an extended gravitationally bound core consisting of five rich Abell clusters, A2056, A2061, A2065, A2067 and A2089. Given the potential importance of these objects, having a number of tools to help identify them is paramount.

The spherical collapse model (SCM) has often been used to study non-linear evolution in the presence of a cosmological constant \citep{Lahav1991,Wang1998,Nagamine2003,Percival2005,Hoffman2007}. Of particular interest are two previous studies that attempted to define the limits of gravitationally bound structure in a universe undergoing accelerated expansion. \citeauthor{Busha03} (\citeyear{Busha03}, hereafter \citetalias{Busha03}) used the SCM, along with the assumption that galaxies would currently be receding from overdensities with an unperturbed Hubble flow velocity, to develop an analytical model for the limits of bound structures in a $\Lambda$CDM universe. However, structures that are destined to be gravitationally bound will have likely slowed from the overall Hubble expansion at present, meaning this model is quite likely to underestimate the extent of bound structure.

\citeauthor{Dunner06} (\citeyear{Dunner06}, hereafter \citetalias{Dunner06}), seeking to improve upon the model of \citetalias{Busha03}, applied a pure SCM to determine the limits of bound structure. Beginning with the Tolman-Bondi equation, and transforming it into dimensionless variables, \citetalias{Dunner06} were able to determine the radius of the `critically bound shell' without assuming anything about the present velocities of galaxies. This model was suggested as being more realistic since it should account for the slowing of `shells' that are part of a bound structure. However, the pure SCM ignores the effects of external attractors which will be present in the real Universe. Since these external attractors will pull on the extremities of a structure, those parts may not actually be bound. This leads the model of \citetalias{Dunner06} to be optimistic in its limits to gravitationally bound structure.

By defining a parameter $\xi$, here named the critical parameter, the results of both models can be expressed as,
\begin{equation}
\label{eq:CritParam}
\dfrac{M_{\mathrm{obj}}}{10^{12} \, \mathrm{M}_{\sun}} \geq \xi \, h_{70}^{2} \left(\dfrac{r_{0}}{1 \, \mathrm{Mpc}}\right)^{3}.
\end{equation}
Here, $M_{\mathrm{obj}}$ is the mass contained within the radius $r_{0}$, and \mbox{$h_{70} = H_{0}/70 \, \mathrm{km} \, \mathrm{s}^{-1} \, \mathrm{Mpc}^{-1}$}, instead of the usual \mbox{$h = H_{0}/100 \, \mathrm{km} \, \mathrm{s}^{-1} \, \mathrm{Mpc}^{-1}$}. Thus, locating the limits of bound structure in the spherical approximation becomes the pursuit of that single parameter. \citetalias{Busha03} found a value of \mbox{$\xi_{\mathrm{B03}} = 3$}, which has a certain appeal due to its simplicity, while \citetalias{Dunner06} found a value of \mbox{$\xi_{\mathrm{D06}} = 1.18$}. In an OCDM universe (\mbox{$\Lambda=0$}), so long as the total energy of a shell, kinetic plus potential, is negative, the shell will remain bound \citep{Nagamine2003}. However, to give some idea as to the sensitivity of $\xi$ to the value of $\Lambda$ we can consider a shell which has a Hubble expansion velocity at present, i.e. the \citetalias{Busha03} model. \cite{Hoffman2007} found that for such a shell to have a turn around in an OCDM universe, the enclosed overdensity must be \mbox{$\delta_{\mathrm{c}} = 2.33$}, which corresponds to a value of $\xi = 0.57$.

In this paper, a more robust estimate of $\xi$ is sought using different methods. First, the results of the $N$-body simulations performed by \citeauthor{Pearson13} (\citeyear{Pearson13}, hereafter \citetalias{Pearson13}) are used to obtain an estimate for $\xi$. Then, a simple modification to the standard SCM is proposed based on an observation of radial and tangential velocities in cosmological simulations. From that modified SCM, an estimate of $\xi$ is made in the same manner as in \citetalias{Dunner06}. The performance of these critical parameters is then tested with cosmological simulations performed with \textsc{gadget2}, and with applications to the CSC and the Virgo cluster.

This paper is structured as follows: Section 2 presents the \citetalias{Pearson13} simulation informed model. Section 3 presents the modified SCM. Section 4 is the comparison with the \textsc{gadget2} simulations. Section 5 demonstrates ways the analytical models may be applied to real structures in our Universe, so long as redshift independent distances are known. Lastly, in Section 6 the implications of the results are discussed.

Throughout this paper it is assumed that \mbox{$\Omega_{\mathrm{m},0} = 0.3$}, \mbox{$\Omega_{\Lambda,0} = 0.7$}, and \mbox{$h = H_{0}/100 \, \mathrm{km} \, \mathrm{s}^{-1} \, \mathrm{Mpc}^{-1}$}.

\section{P13 Simulation Informed Limits}
The simulations of \citetalias{Pearson13} imply that the model of \citetalias{Busha03} would act as a lower limit to the true extent of bound structure, while the model of \citetalias{Dunner06} would act as an upper limit. Given the excellent agreement of the SSC simulation results with the results of \cite{Munoz2008} as to the extent of bound structure, it is reasonable to assume that the simulation results are reliable.

The simulations of the CSC performed by \citetalias{Pearson13} showed very little chance of there being extended gravitationally bound structure, due to using spectroscopic redshifts as distance indicators when the clusters had significant peculiar motions along the line-of-sight (\citeauthor{Batiste13} \citeyear{Batiste13}; \citetalias{Pearson14}). However, there were some cases in which two different pairs of clusters showed up as bound, A2061/A2067 and A2065/A2089. It seems safe to then assume that in each of these cases, they must have been near the limits of bound structure. For this reason, they were examined to determine a value of $\xi$. Before that was done, the further requirement that the current kinetic energy of the pair be at least 95 per cent of the potential energy was applied, ensuring that the pairs were loosely bound.

\begin{table}
\centering
\parbox{\linewidth}{
\caption{CSC Critical Parameter Constraints. Column 1 lists the cluster pair being examined. Column 2 lists the mass of the larger cluster in the pair. Column 3 lists the initial separation, and Column 4 is the value of the critical parameter.}
\label{CBSC2}
\centering
\begin{tabular}{@{}lccc@{}}
\toprule
Cluster Pair & $M_{\mathrm{obj}}$ & $r_{0}$ & $\xi$ \\
${}$ & ($10^{15} \, h^{-1} \, \mathrm{M}_{\odot}$) & ($h^{-1} \, \mathrm{kpc}$) & ${}$ \\
\midrule
A2061/A2067 & 1.708$^{a}$ & 6326.80 & 2.96 \\
${}$ & ${}$ & 8775.97 & 1.11 \\
${}$ & ${}$ & 8417.99 & 1.26 \\
${}$ & ${}$ & 6548.20 & 2.67 \\
${}$ & ${}$ & 7426.02 & 1.83 \\
${}$ & ${}$ & 9941.05 & 0.76 \\
${}$ & ${}$ & 7039.69 & 2.15 \\
${}$ & ${}$ & 6162.42 & 3.20 \\
${}$ & ${}$ & 7928.76 & 1.50 \\
${}$ & ${}$ & 7449.05 & 1.81 \\
${}$ & ${}$ & 8173.69 & 1.37 \\
${}$ & ${}$ & 8102.29 & 1.41 \\ [1ex]
A2065/A2089 & 0.951$^{a}$ & 5690.67 & 2.26 \\
${}$ & ${}$ & 5044.48 & 3.25 \\
\bottomrule
\end{tabular}
\\ [1ex]
\raggedright
$^{a}$Mass is listed once for reference. All values for a given pair use the same mass to determine $\xi$.
}
\end{table}

By assuming that the cluster pairs were at the limits of bound structure, the equality in equation \eqref{eq:CritParam} was taken and then solved for the critical parameter,
\begin{equation}
\label{eq:CritParamCalc}
\xi = \dfrac{M_{\mathrm{obj}}}{h_{70}^{2}(10^{12} \, \mathrm{M}_{\sun})}\left(\dfrac{1 \, \mathrm{Mpc}}{r_{0}}\right)^{3}.
\end{equation}
The initial separations and mass of the larger of the two clusters can be entered into equation \eqref{eq:CritParamCalc}, giving a value for the critical parameter. The results of applying this procedure to the CSC cluster pairs are shown in Table \ref{CBSC2}. Looking at the values of the critical parameters, we can see that there are two cases in which the criterion is more conservative than that of \citetalias{Busha03}, meaning those are cases where the structure should definitely be bound. There are also two cases where the criterion is lower than that of \citetalias{Dunner06}, implying the clusters must have had motions towards each other, leading them to be bound. The rest lie somewhere between those two models. Taking a bi-weight median \citep{Beers90}, a value of \mbox{$\xi = 1.90$} is obtained. Confidence limits are determined using jackknife resampling \citep{Beers90}, giving in the end
\mbox{$\xi_{\mathrm{sim}} = 1.90 \pm 0.37$} with 95 per cent confidence. This confirms more rigorously the findings of \citetalias{Pearson13}, that the true limits to bound structure, at least in those simulations, lie between the \citetalias{Busha03} and \citetalias{Dunner06} models.

\section{A Modified SCM}
\begin{figure}
\includegraphics[width=1.0\linewidth]{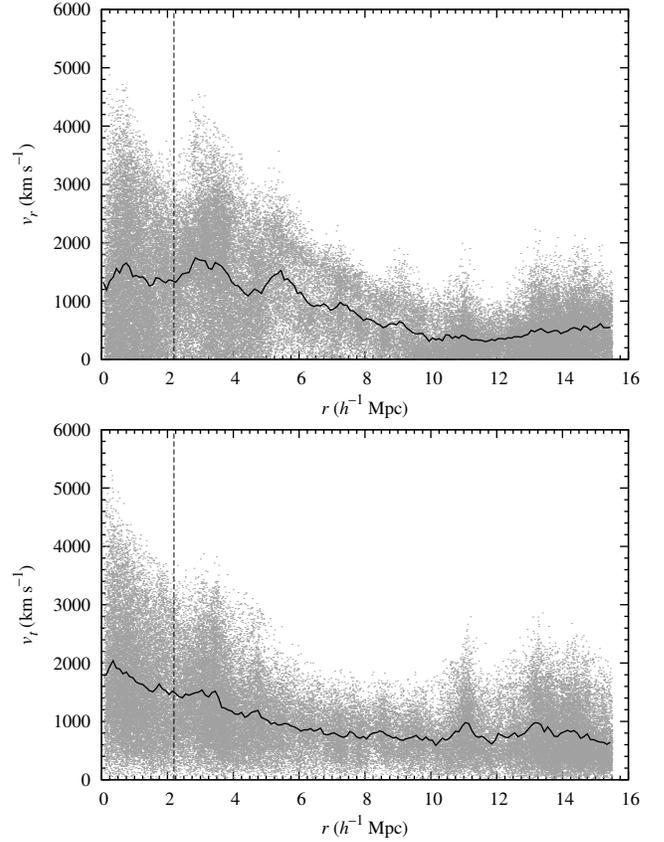}
\caption{Radial and tangential velocities of particles in Object 1 from Simulation 2 as a function of radius in the top and bottom panels, respectively. The magnitudes of the radial velocities are plotted. The vertical dashed lines in both panels mark $r_{200}$ (density inside is \mbox{$200\rho_{\mathrm{c},0}$}) for this object. The solid lines show average velocities in \mbox{$100 \, h^{-1} \, \mathrm{kpc}$} bins. Note the significant tangential motions of particles well outside of the virialized region.}
\label{fig:VrvsVtan}
\end{figure}

When it comes to applying the SCM to structures in the Universe, aside from the implicit assumption of spherical symmetry, there are two key weaknesses. First, the SCM excludes the effects of what can be called external attractors, mass concentrations not bound to the central dominant cluster, which can pull galaxies away from the structure. Second, the SCM assumes that all motions are purely radial. A quick examination of particles from simulation shows that there will be some fairly substantial tangential motions at present around structures that are still in the formation process. Fig. \ref{fig:VrvsVtan} shows the radial and tangential motions of particles around Object 1 from Simulation 2 (see section \ref{sec:GadgetComp}) at the present time (\mbox{$a=1$}). The bottom panel shows that there are significant tangential motions all the way out to the turn around radius (\mbox{$\sim 11 \, h^{-1} \, \mathrm{Mpc}$}). Comparing with the top panel, we can see that it is reasonable to assume equal radial and tangential motions.

In the standard SCM, assuming only radial motions, the energy equation is \citep{Peebles1980}
\begin{equation}
\label{eq:TolmanBondi}
E = \dfrac{1}{2}v_{r}^{2}-\dfrac{GM}{r} - \dfrac{1}{6}\Lambda r^{2},
\end{equation}
where $E$ is the energy, $v_{r}$ is the radial velocity, $G$ is the gravitational constant, $M$ is the enclosed mass, $r$ is the radius of the shell, and $\Lambda$ is the cosmological constant. Adding in some tangential motions and making the assumption that they are roughly equal to the radial motions, we arrive at
\begin{equation}
\label{eq:SCMMod2}
E = v_{r}^{2}-\dfrac{GM}{r} - \dfrac{1}{6}\Lambda r^{2}.
\end{equation}

Following the prescription of \citetalias{Dunner06}, equation \eqref{eq:SCMMod2} is transformed into dimensionless variables, defined as:
\begin{equation}
\widetilde{r} = \left(\dfrac{\Lambda}{3GM}\right)^{1/3}r,
\end{equation}
\begin{equation}
\widetilde{t} = \left(\dfrac{\Lambda}{3}\right)^{1/2}t,
\end{equation}
\begin{equation}
\widetilde{v}_{r} = \left(\dfrac{3}{G^{2}M^{2}\Lambda}\right)^{1/6} v_{r}
\end{equation}
The energy equation is then rearranged to solve for the dimensionless time, making use of the fact that \mbox{$\widetilde{v}_{r} = d\widetilde{r}/d\widetilde{t}$}, giving
\begin{equation}
\label{int2d}
\widetilde{t}_{0} = \sqrt{2} \int_{0}^{\widetilde{r}_{0}} \left(\dfrac{\widetilde{r}}{\widetilde{r}^{3} + 2\widetilde{E} \widetilde{r} + 2}\right)^{1/2} d\widetilde{r}.
\end{equation}
The difference between equation \eqref{int2d} and the similar equation from \citetalias{Dunner06} is a multiplying factor of $\sqrt{2}$. This integral can be evaluated analytically for the critical energy (\mbox{$\widetilde{E} = -3/2$}, see \citetalias{Dunner06}). After performing the integration, it is useful to define a new variable $\chi(\widetilde{r}_{\mathrm{cs}})$, as done in \citetalias{Dunner06}, which will have a different exponent than they found,
\begin{equation}
\label{chi2d}
\begin{array}{ll}
\chi(\widetilde{r}_{\mathrm{cs}}) = & \left[\dfrac{1 + 2\widetilde{r}_{\mathrm{cs}} + \sqrt{3\widetilde{r}_{\mathrm{cs}}(\widetilde{r}_{\mathrm{cs}}+2)}}{1 + 2\widetilde{r}_{\mathrm{cs}} - \sqrt{3\widetilde{r}_{\mathrm{cs}}(\widetilde{r}_{\mathrm{cs}}+2)}}\right]^{\sqrt{3/2}} \\
& \times \left(1 + \widetilde{r}_{\mathrm{cs}} + \sqrt{\widetilde{r}_{\mathrm{cs}}(\widetilde{r}_{\mathrm{cs}}+2)}\right)^{-3\sqrt{2}}.
\end{array}
\end{equation}
Using the fact that
\begin{equation}
\Omega_{\Lambda} \equiv \dfrac{\Lambda}{3H^{2}} = \tanh^{2}\left(\dfrac{3\widetilde{t}}{2}\right),
\end{equation}
this variable is then related to \mbox{$\Omega_{\Lambda}$} via
\begin{equation}
\label{OmegaL}
\Omega_{\Lambda}(\widetilde{r}_{\mathrm{cs}}) = \left[\dfrac{\chi(\widetilde{r}_{\mathrm{cs}})-1}{\chi(\widetilde{r}_{\mathrm{cs}})+1}\right]^{2}.
\end{equation}
This expression was numerically evaluated by incrementally increasing $\widetilde{r}_{\mathrm{cs}}$, starting from zero, until \mbox{$\Omega_{\Lambda}(\widetilde{r}_{\mathrm{cs}}) = 0.7$}. This occured for \mbox{$\widetilde{r}_{\mathrm{cs}} = 0.75$}. Framing this in terms of a critical parameter, as in equation \eqref{eq:CritParam}, yields \mbox{$\xi_{\mathrm{P14}} = 1.89$} in excellent agreement with \mbox{$\xi_{\mathrm{sim}}$}. This is hereafter referred to as the P14 model.

\begin{figure}
\includegraphics[width=1.0\linewidth]{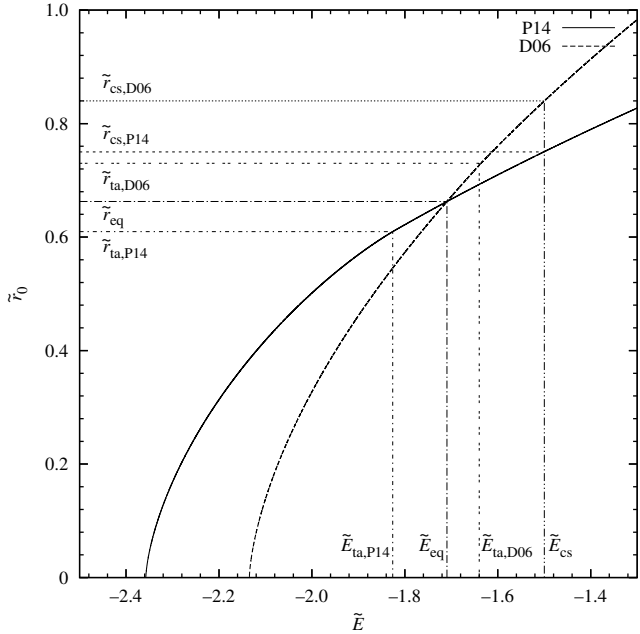}
\caption{A comparison of the \citetalias{Dunner06}, based on the pure SCM, and the P14 models. The critical and turn around radii of each model are labeled, along with the point at which the two models have equivalent results. Noting the locations of the turn around radii, at $\widetilde{r}_{\mathrm{eq}}$ the \citetalias{Dunner06} model would have that shell collapsing, while the P14 model would have it still expanding.}
\label{fig:D06vsP14}
\end{figure}

Equation \eqref{int2d} was also numerically integrated for energies other than the critical energy. An energy value was selected, then the right hand side was numerically integrated until it equaled the value of the current dimensionless time, $\widetilde{t}_{0}$. This procedure was performed for both the P14 and \citetalias{Dunner06} models, starting with \mbox{$\widetilde{E} = 0$}, incrementing until a value of energy where the current radius would be zero. The results of these integrations are shown in Fig. \ref{fig:D06vsP14}. There are several noteworthy features from this figure. The turn around radius of the \citetalias{Dunner06} model (\mbox{$\widetilde{r}_{\mathrm{ta,D06}}=0.73$}) is only slightly smaller than the critical radius of the P14 model (\mbox{$\widetilde{r}_{\mathrm{cs,P14}}=0.75$}), and is significantly larger than the turn around radius of the P14 model (\mbox{$\widetilde{r}_{\mathrm{ta,P14}}=0.61$}). There is also a point at which the two models cross (\mbox{$\widetilde{r}_{\mathrm{eq}} = 0.66$}) which lies between the two turn-around radii, meaning the P14 model would have that shell expanding, while that of \citetalias{Dunner06} would have it collapsing. These features can potentially be used to test the performance of the models when compared to simulation data.

In order to carry out a comparison, the densities associated with these shells need to be known, so that they can be located in simulation data. The density parameter for any shell is given by \citepalias[equation (17)]{Dunner06}
\begin{equation}
\label{eq:ShellDenParam}
\Omega_{\mathrm{s}} \equiv \dfrac{\bar{\rho}_{\mathrm{m}}^{\mathrm{s}}}{\rho_{\mathrm{c}}} = \dfrac{2\Omega_{\Lambda}}{\widetilde{r}^{3}}.
\end{equation}
Since the dimensionless radii of all the shells of interest are known, the calculation of the density parameters is trivial. The density needed to locate the limits of bound structure in the future can also be located easily by using \mbox{$\Omega_{\mathrm{cs}}(a=100) = 2$} \citepalias{Dunner06}, and multiplying by the future (\mbox{$a=100$}) value of the critical density. The value for $\widetilde{r}_{\mathrm{cs,B03}}$ is determined from the parameter $\beta$ of \citetalias{Busha03} defined as
\begin{equation}
\label{b03beta}
\beta \equiv \dfrac{2GM}{H_{0}r_{0}^{3}}.
\end{equation}
While they name this the strength parameter, it is actually just the shell density parameter, which we can see by rearranging equation \eqref{b03beta} as
\begin{equation}
\beta = \dfrac{3M}{4\mathrm{\pi}r_{0}^{3}}\dfrac{8\mathrm{\pi}G}{3H_{0}^{2}}.
\end{equation}
The first part of this equation is simply the enclosed mass divided by the volume, or the mean mass density, and the second part is the inverse of the critical density. From their analysis, \citetalias{Busha03} determine that in order for a shell to have a turn around, $\beta$ must satisfy
\begin{equation}
\label{b03critical}
\beta^{3} \geq \dfrac{27}{4}(\Omega_{\mathrm{m,0}}+\beta)^{2}\Omega_{\Lambda,0}.
\end{equation}
Evaluation equation \eqref{b03critical} at the equality gives the critical density parameter for their model, allowing for the determination of the density and associated dimensionless radius. The results are shown in Table \ref{tab:ShellDens}.

\begin{table}
\centering
\parbox{\linewidth}{
\caption{Densities associated with various radii from the \citetalias{Busha03}, P14, and \citetalias{Dunner06} models. Column 1 labels which radius is being considered. Column 2 gives the value of that radius in dimensionless units. Column 3 gives the shell density parameter. Lastly, column 4 gives the shell density. The units of density are chosen to match the \textsc{gadget2} simulation units.}
\label{tab:ShellDens}
\centering
\begin{tabular}{@{}lccc@{}}
\toprule
Radius & $\widetilde{r}$ & $\Omega_{\mathrm{s}}$ & $\rho_{\mathrm{s}}$ \\
${}$ & ${}$ & ${}$ & ($10^{10} \, h^{2} \, \mathrm{M}_{\sun} \, \mathrm{kpc}^{-3}$) \\
\midrule
$\widetilde{r}_{\mathrm{cs}}(a=100)$ & 1 & 2 & $3.88 \times 10^{-8}$ \\
$\widetilde{r}_{\mathrm{cs,D06}}$ & 0.84 & 2.36 & $6.55 \times 10^{-8}$ \\
$\widetilde{r}_{\mathrm{cs,P14}}$ & 0.75 & 3.32 & $9.21 \times 10^{-8}$ \\
$\widetilde{r}_{\mathrm{ta,D06}}$ & 0.73 & 3.60 & $9.99 \times 10^{-8}$ \\
$\widetilde{r}_{\mathrm{eq}}$ & 0.66 & 4.87 & $1.35 \times 10^{-7}$ \\
$\widetilde{r}_{\mathrm{cs,B03}}$ & 0.64 & 5.28 & $1.48 \times 10^{-7}$ \\
$\widetilde{r}_{\mathrm{ta,P14}}$ & 0.61 & 6.17 & $1.71 \times 10^{-7}$ \\
\bottomrule
\end{tabular}
}
\end{table}

\section{Comparisons with \textsc{gadget2} Simulation}
\label{sec:GadgetComp}
In order to rigorously compare all three models, two simulations were performed with \textsc{gadget2} \citep{Springel05}. Simulation 1 was designed to reproduce that performed by \citetalias{Dunner06}: $128^{3}$ particles of mass \mbox{$3.97 \times 10^{10} \, h^{-1} \, \mathrm{M}_{\sun}$} in a \mbox{$100 \, h^{-1} \, \mathrm{Mpc}$} periodic box, \mbox{$\Omega_{\mathrm{m,0}} = 0.3$}, \mbox{$\Omega_{\Lambda,0} = 0.7$}, \mbox{$\sigma_{8} = 1$}, and the softening length was set to \mbox{$15 \, h^{-1} \mathrm{kpc}$} in co-moving coordinates until the present time and then capped at that value in physical units. Simulation 2 was larger: \mbox{$256^{3}$} particles of mass \mbox{$13.4 \times 10^{10} \, h^{-1} \, \mathrm{M}_{\sun}$} in a \mbox{$300 \, h^{-1} \, \mathrm{Mpc}$} periodic box, \mbox{$\Omega_{\mathrm{m,0}} = 0.3$}, \mbox{$\Omega_{\Lambda,0} = 0.7$}, \mbox{$\sigma_{8} = 1$}, and the softening length was set to \mbox{$50 \, h^{-1} \mathrm{kpc}$} in co-moving coordinates until the present time, after which it was capped at that value in physical units. Both simulations were evolved from \mbox{$a = 0.02$} (\mbox{$z = 49$}) to \mbox{$a = 100$}. Initial conditions were generated using the \textsc{n-g}en\textsc{ic} software, also developed by Volker Springel. The purpose of Simulation 1 was to check the methods used here for consistency with those of \citetalias{Dunner06}. However, given the relatively small box size of that simulation combined with the periodic boundary conditions, the long range gravitational effects may have been somewhat suppressed. These forces might be very influential in terms of the effects of external attractors on potentially gravitationally bound structures. The larger box size of Simulation 2 should have allowed for more realistic long range gravitational forces, better modelling the effects of external attractors.

Both simulations were analyzed in the same way. Snapshots were taken at the present (\mbox{$a = 1$}) and in the far future when structure formation should decrease significantly (\mbox{$a = 100$}). Structures were identified using the new software  \textsc{coldg}a\textsc{s}\footnote{For this work, \textsc{coldg}a\textsc{s}-0.4.3 was used and is publicly available at \href{http://davidwpearson.com/some-of-my-code/}{http://davidwpearson.com/some-of-my-code/}}, Compute Unified Device Architecture (CUDA) Object Location Determination for \textsc{gadget2} Snapshots, developed by the author. As the title suggests, the code uses NVidia's CUDA language to leverage the massively parallel capabilities of graphics processor units (GPUs) to significantly speed up certain portions. The code first performs a grid based density calculation in which particles are assigned to cells by dividing each coordinate of the particles position by the length of one side of a cell, and rounding the result down. This gives three indices corresponding to a unique cell. In this manner, the density calculation will always have a complexity of order $4N$, regardless of the size of the grid. The particles are sorted according to which cell they belong in, so that only information about occupied cells is stored, reducing memory costs for small grid sizes. Density peaks are then used to find galaxy cluster like objects, the most massive of which go on to become the centres of structures to be studied as potentially gravitationally bound. After some further improvements are made, a future paper will provide a much more detailed description of the software.

Both the present (\mbox{$a=1$}) and future (\mbox{$a=100$}) snapshots were analyzed with \textsc{coldg}a\textsc{s}. The most massive future objects were matched to their progenitors at present based on their positions in the simulation volume. Only objects that were sufficiently far away from the edges so that their radii would not overlap the box boundary were chosen for simplicity. Once objects were matched between the present and future snapshots, particle IDs were compared to determine which particles remained part of the structure into the far future, implying that they are gravitationally bound.

\subsection{Simulation 1}
The primary reason for Simulation 1 was to verify that the methods used here were consistent with those used by \citetalias{Dunner06}. In order to check this, the output of that simulation was analyzed in the same way as the data of \citetalias{Dunner06}. Using the above described procedure, 20 objects were selected from this simulation. Their future masses ranged between $2.33$ and \mbox{$19.0 \times 10^{14} \, h^{-1} \, \mathrm{M}_{\sun}$}. The range in masses at present is dependent upon the chosen analytical model, but is in rough agreement with the above limits.

Table \ref{tab:ModelComp} shows a summary of the results from analyzing the data in the same manner as was done in \citetalias{Dunner06}, with the results on a per object basis available as online supplemental material. The four quantities presented are the same as calculated by \citetalias{Dunner06} in which the objects are examined on a per particle basis. The first two rows of Table \ref{tab:ModelComp} show the results obtained by \citetalias{Dunner06} when they analyzed their simulation. Comparing those first two rows with the results obtained here in applying the models of \citetalias{Busha03} and \citetalias{Dunner06} to Simulation 1, we can see that they agree within the uncertainties. This implies that the methods of analysis used here are consistent with those used by \citetalias{Dunner06}. As expected, the results of the P14 model fall between the \citetalias{Busha03} and \citetalias{Dunner06} models. Both the \citetalias{Busha03} and P14 models do well at finding the final bound mass of the objects, while the model of \citetalias{Dunner06} predicts far more mass being bound. However, all of the model's results agree with each other at the $2\sigma$ level, indicating that this is not the most informative way of examining their performance.

\begin{table}
\centering
\parbox{\linewidth}{
\caption{Summary of the analysis of Simulation 1 in the manner of \citetalias{Dunner06}. Column 1 identifies the model. Column 2 is the ratio of particles identified as bound at present that escape to the total number selected as bound at present. Column 3 is the ratio of particles identified as bound at present that remain part of the structure to the total number selected as bound at present. Column 4 is the ratio of particles that are part of the structure in the future but were not selected as bound at present to the total number selected as bound at present. Lastly, Column 5 is the ratio of mass that is bound in the future to the mass selected as bound in the present.}
\label{tab:ModelComp}
\centering
\begin{tabular}{@{}lr@{$\: \pm \:$}lr@{$\: \pm \:$}lr@{$\: \pm \:$}lr@{$\: \pm \:$}l@{}}
\toprule										
Model	&	\multicolumn{2}{c}{A}	&	\multicolumn{2}{c}{B}	&	\multicolumn{2}{c}{C}	&	\multicolumn{2}{c}{D} \\
${}$	&	\multicolumn{2}{c}{Per cent}	&	\multicolumn{2}{c}{Per cent}	&	\multicolumn{2}{c}{Per cent}	&	\multicolumn{2}{c}{Per cent} \\
\midrule
B03$^{a}$ &	9.9 & 4.2 & 90.1 & 4.2 & 13.3 & 12.4 & 103.4 & 14.3 \\
D06$^{b}$ & 28.2 & 13.0 & 71.8 & 13.0 & 0.26 & 0.23 & 72.0 & 13.1 \\ [1ex]
B03	&	8.7 & 4.0	&	91.3 & 4.0 &	20.0 & 22.0	&	111.3 & 19.4 \\
D06	&	19.5 & 7.0	&	80.5 & 7.0	&	1.6 & 2.5	&	82.0 & 7.0	\\ 
P14	&	13.7 & 6.0	&	86.3 & 6.0	&	6.9 & 10.2	&	93.1 & 10.1	\\ 
\bottomrule											
\end{tabular}
\\ [1ex]
\raggedright
$^{a}$Data in this row is from Table 2 of \citetalias{Dunner06}.\\
$^{b}$Data in this row is from Table 1 of \citetalias{Dunner06}.\\
}
\end{table} 

For all of the analytical models, the dimensionless critical radius should tell us the relative size of the present critical shell compared to its final critical radius, since all models should approach a dimensionless radius of \mbox{$\widetilde{r} = 1$} as \mbox{$t \rightarrow \infty$}. For this reason a comparison was made between the present and future critical radii, to see if any one model performed better. For the model of \citetalias{Busha03} the critically bound shell should be about 64 per cent of its final size, and in Simulation 1 it is on average \mbox{$62.6 \pm 3.4$} per cent of its final size. For the P14 model, the critically bound shell should be about 75 per cent of its final size, and in Simulation 1 it is on average \mbox{$77.3 \pm 2.8$} per cent of its final size. Lastly, the model of \citetalias{Dunner06} predicts the critically bound shell to be about 84 per cent of its final size, while in Simulation 1 it is found to be, on average, \mbox{$90.2 \pm 2.9$} per cent of its final size. Taking only the central values into account would suggest that the model of \citetalias{Busha03} is slightly conservative, the P14 model is slightly optimistic, and the \citetalias{Dunner06} model is much more optimistic. However, for all of the models, the predicted ratio is within $2\sigma$ of the average ratio found in Simulation 1, meaning that we cannot learn much from this analysis.

It is well known that when two structures merge, some of the mass of those structures is ejected \citep{White1978,Colpi1999}. Combining this with the fact that all models agree with each other at the $2\sigma$ level, strongly suggests that trying to examine structures from simulation on a per particle basis does not make much sense. What is of more interest is being able to identify which clusters may be gravitationally bound to each other, and which are not. This implies that a more useful way of assessing the models' performance would be to visually examine the objects, to see which model does the best at selecting the clusters that are bound to the central dominant cluster. By using the IDs of particles that are part of the object in the future snapshot, it is possible to pull the coordinates of those particles from the present snapshot data. Plotting these particles will then give a good idea of what is bound to the central dominant cluster. Because of mass being lost during mergers, the plots containing only the future particles will seem less dense than the plots from the analytical models.

\begin{figure*}
\includegraphics[width=0.49\linewidth]{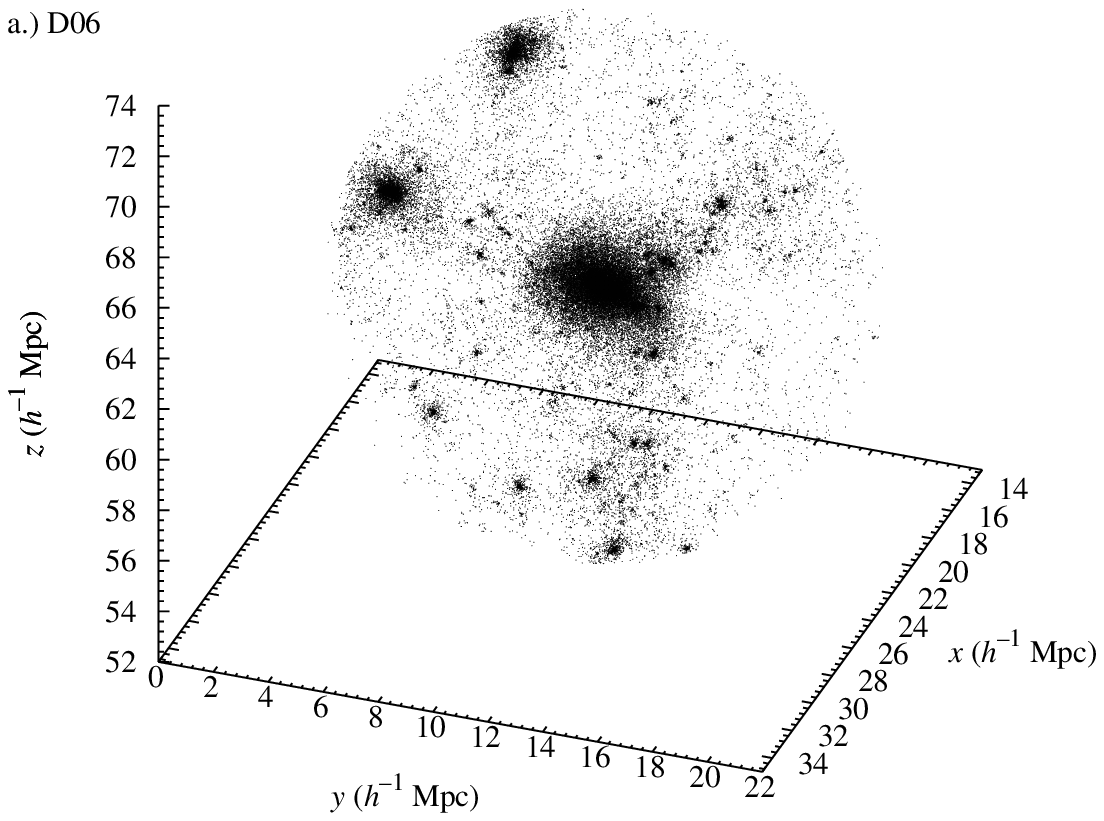} \hfill
\includegraphics[width=0.49\linewidth]{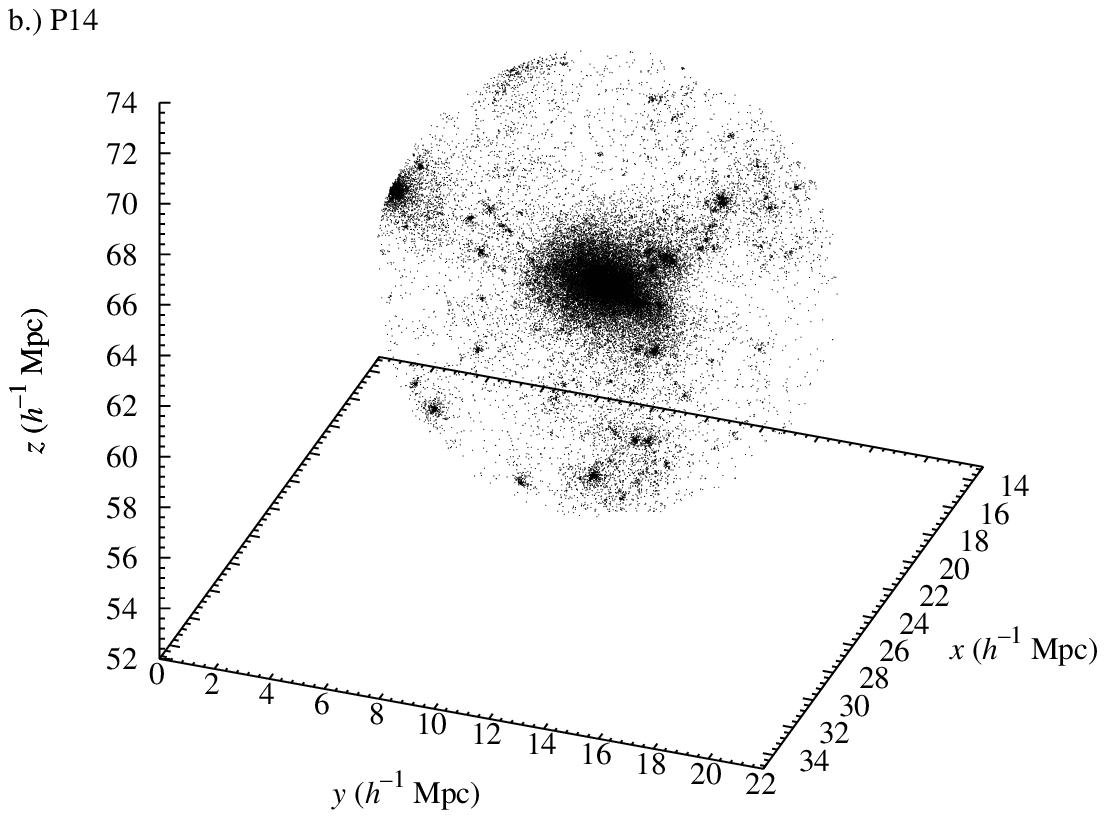}

\includegraphics[width=0.49\linewidth]{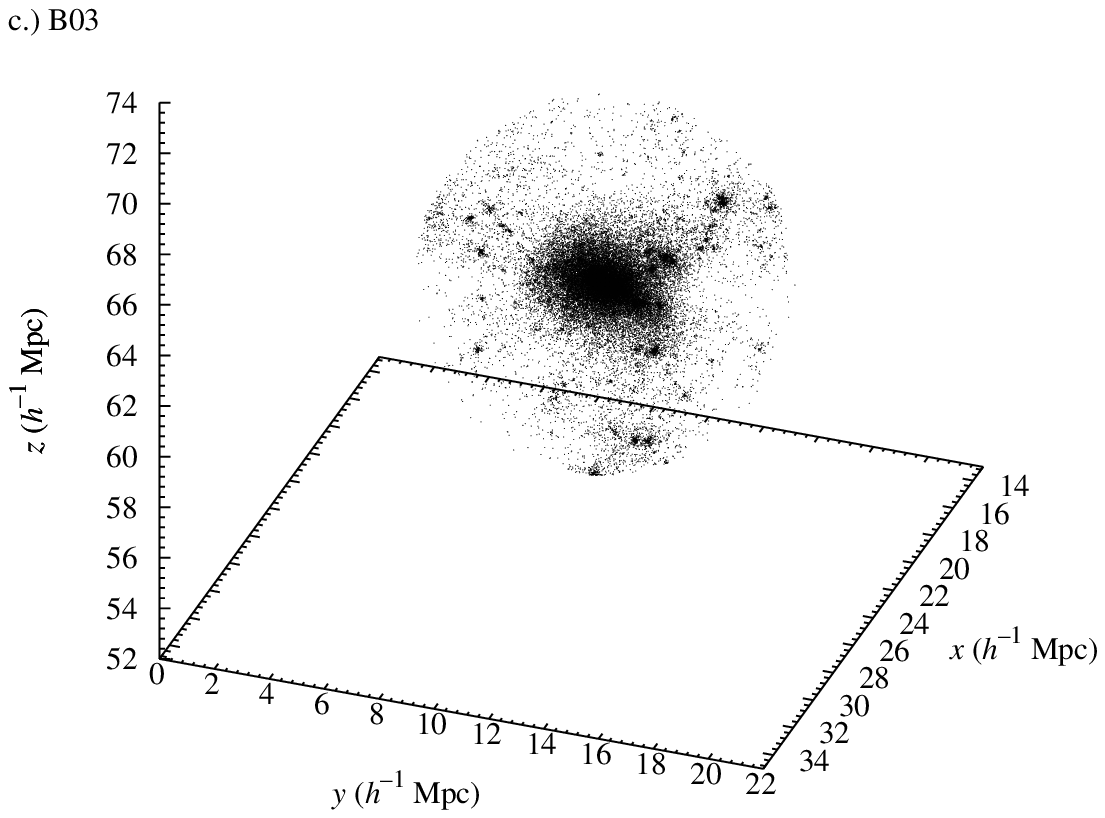} \hfill
\includegraphics[width=0.49\linewidth]{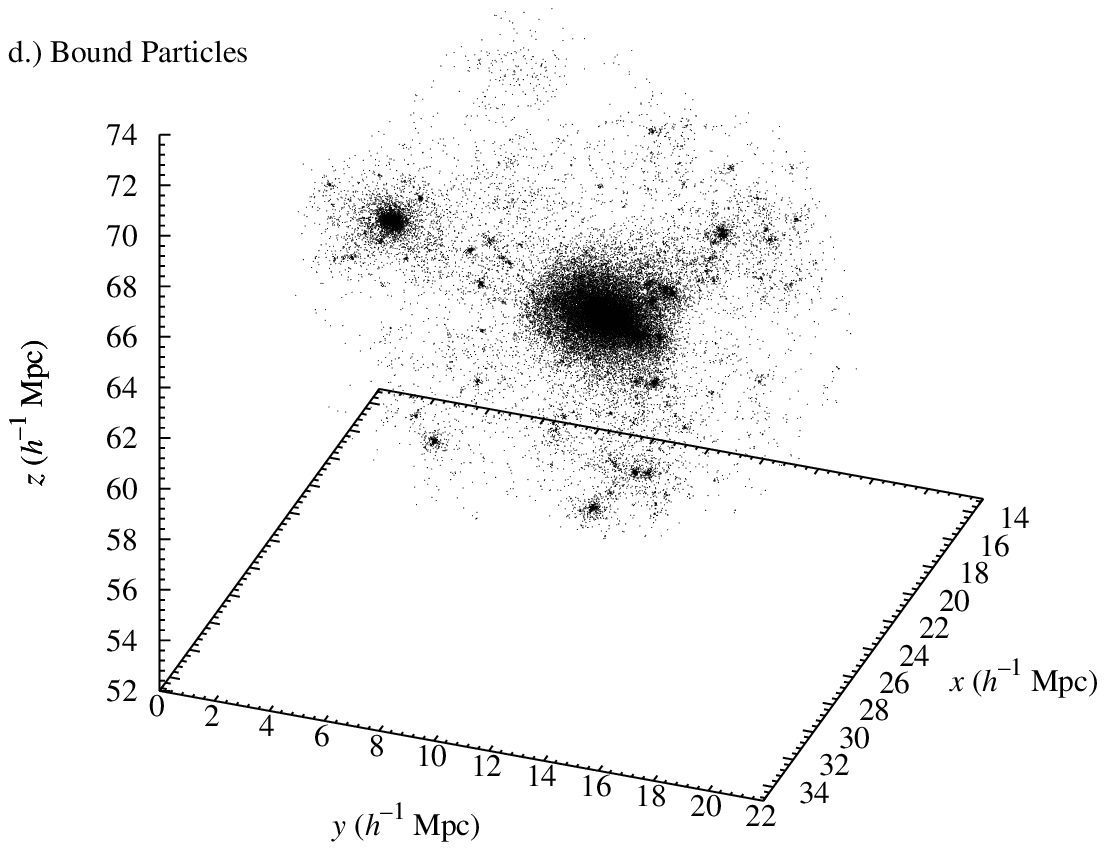}

\includegraphics[width=0.49\linewidth]{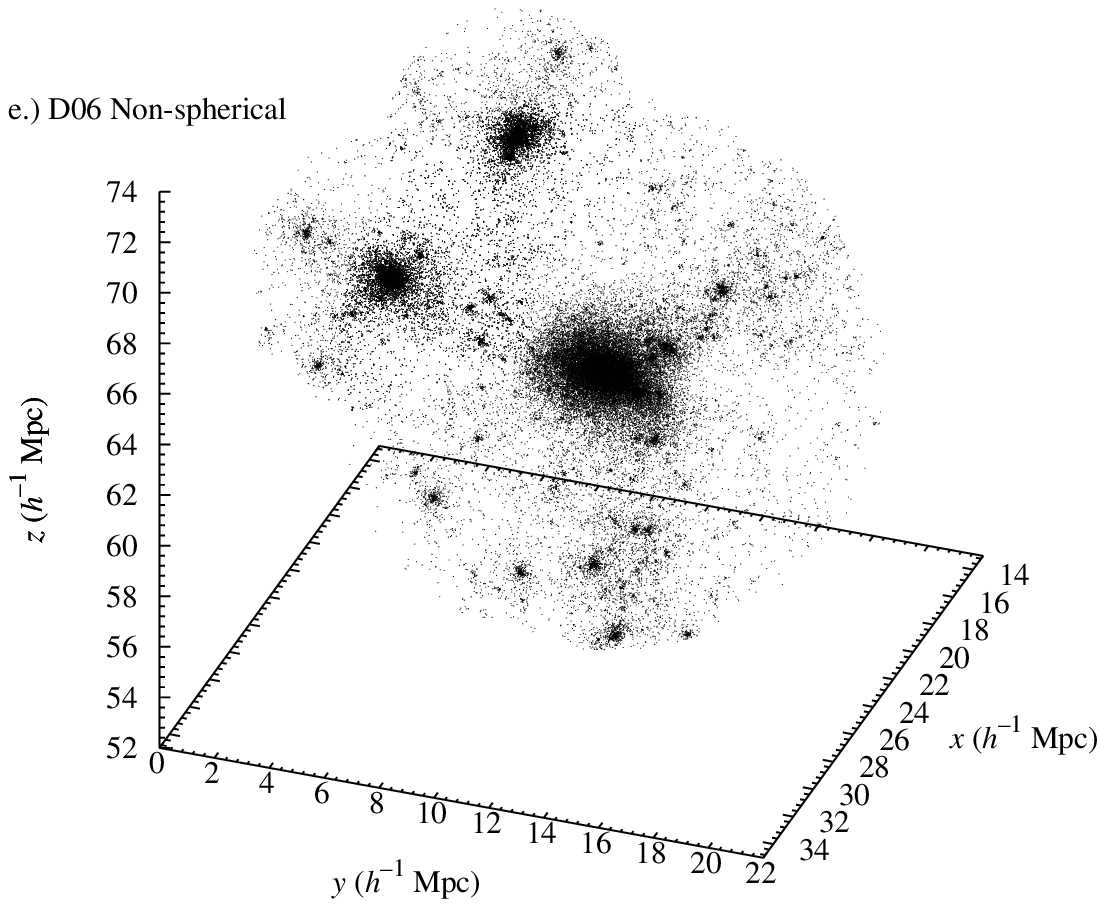} \hfill
\includegraphics[width=0.49\linewidth]{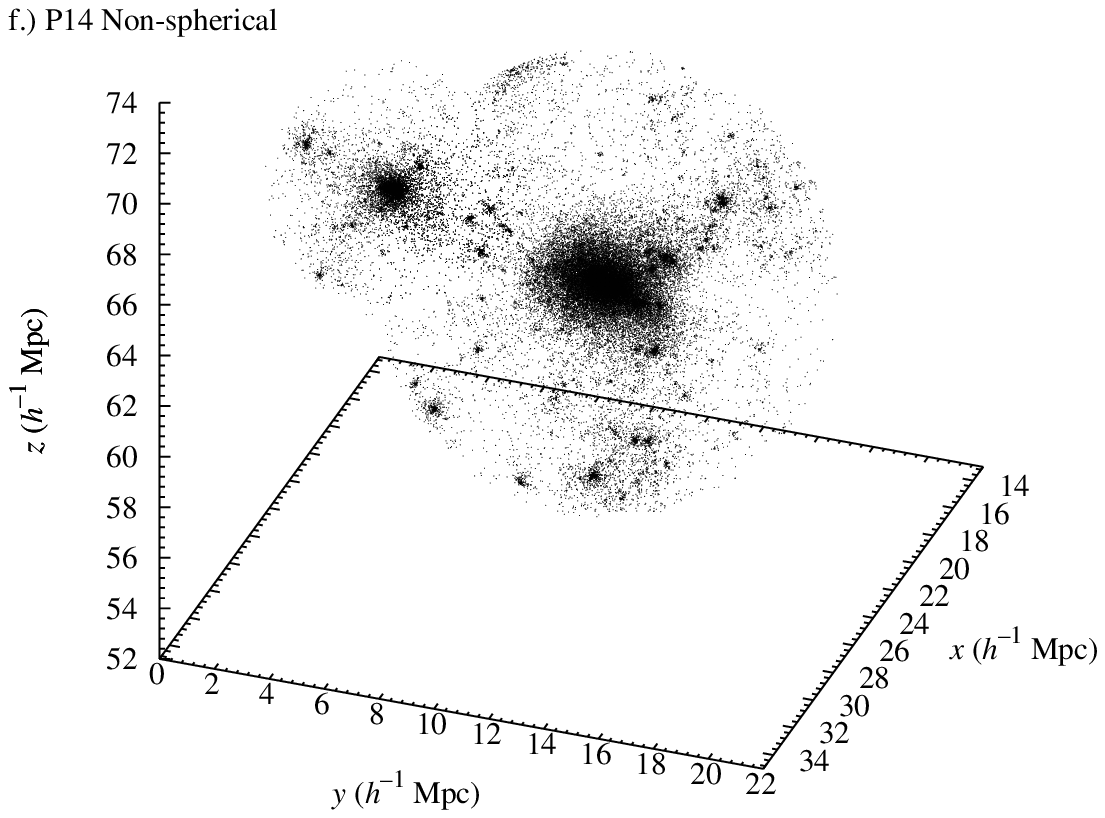}
\caption{Three-dimensional plots of the particles associated with Object 1 from Simulation 1. a.) All particles pulled from the present ($a=1$) snapshot based on the \citetalias{Dunner06} cutoff density. b.) All particles pulled from the present snapshot based on the P14 cutoff density. c.) All particles pulled from the present snapshot based on the \citetalias{Busha03} cutoff density. d.) All particles that are part of the structure in the future ($a=100$) snapshot in their present ($a=1$) locations. e.) All particles pulled from the present snapshot based on the \citetalias{Dunner06} cutoff density applied to the central cluster and two additional clusters. f.) All particles pulled from the present snapshot based on the P14 cutoff density applied to the central cluster and the lower of the two additional clusters. Comparing panels d and f shows that the model of P14 applied in the `non-spherical' manner does well at locating what becomes part of the final structure (i.e. what is gravitationally bound).}
\label{fig:D06SimObj1}
\end{figure*}

Fig. \ref{fig:D06SimObj1} shows the particles selected as potentially bound by the three analytical models applied purely spherically (only from the central cluster), and the particles that are part of the structure in the future frame, excluding particles ejected due to mergers. Also included are plots of the particles selected as potentially bound by the models of \citetalias{Dunner06} and P14 applied in a `non-spherical' manner described below. Comparing the various panels to the panel containing the particles that are part of the final structure (panel d), we can get a good idea about the performance of the analytical models. In the purely spherical mode, the model of \citetalias{Dunner06} identifies two substantial clusters as bound to the central cluster, and the P14 model only includes one of these (the cluster's centre of mass is inside the P14 critical radius by \mbox{$\sim 6 \, h^{-1} \, \mathrm{kpc}$}), while the model of \citetalias{Busha03} does not include either of them. The one just inside the critical radius of the P14 model ends up as bound to the structure, while the other cluster does not. The last two panels of the figure show the `non-spherical' application of the \citetalias{Dunner06} and P14 models. Here the predictions are made by first applying an analytical model to a central dominant cluster, and then to any other clusters within the central clusters critical radius. Note that panel f (the `non-spherical' P14 model) seems to most closely approximate what becomes part of the final structure.

Visual examination of the other objects from Simulation 1 fail to yield any additional insights to the models. For figures similar to Fig. \ref{fig:D06SimObj1} for all of the objects from this simulation, please see the online supplemental material.

So far, the results have been rather unsurprising. The \citetalias{Dunner06} model seems to be optimistic as to the extent of bound structure, while the \citetalias{Busha03} model seems conservative to the point of missing parts of the structure that are bound. The P14 model falls between the two, and detailed examination of the objects hints that it is slightly on the conservative side, though it seems significantly less likely to miss bound portions of the structure than the \citetalias{Busha03} model.

\subsection{Simulation 2}
The larger box size of this simulation should ensure that it better approximated the effects of long-range gravitational forces, which may have been underestimated by Simulation 1. The analysis of this simulation was undertaken in the same manner as for Simulation 1, again selecting 20 objects for study with masses in the future ranging between $1.83$ and \mbox{$7.67 \times 10^{15} \, h^{-1} \, \mathrm{M}_{\sun}$}. As before, the mass range at present will vary depending on the choice of analytical model, but it is noteworthy that the most massive object in this simulation is \mbox{$\sim 10^{16} \, h^{-1} \, \mathrm{M}_{\sun}$} at present according to the P14 and \citetalias{Dunner06} models, similar to the estimated masses of the SSC and CSC.

Table \ref{tab:ModelComp2} shows the results of analyzing the data as was done in \citetalias{Dunner06}. From this table, it is clear that the results are virtually identical between the two simulations. Again we see that the results of applying the \citetalias{Busha03} and \citetalias{Dunner06} models to the simulation performed by \citetalias{Dunner06}, agree with the results of applying those models to Simulation 2. The results of the P14 model fall between the other two, and agree with the results of that model from Simulation 1. The ratios of present radii to final radii are \mbox{$62.6\pm 2.6$}, \mbox{$77.1\pm 2.1$}, and \mbox{$89.2\pm 2.4$} per cent, for the \citetalias{Busha03}, P14, and \citetalias{Dunner06} models, respectively. These again agree with the theoretical expectations at the $2\sigma$ level. For the statistics tracked by \citetalias{Dunner06} all of the models agree with each other at the $2\sigma$ level. Results on a per object basis are provided in an online supplement.

\begin{table}
\centering
\parbox{\linewidth}{
\caption{Summary of the analysis of Simulation 2 in the manner of \citetalias{Dunner06}. Columns are the same as in Table \ref{tab:ModelComp}.}
\label{tab:ModelComp2}
\centering
\begin{tabular}{@{}lr@{$\: \pm \:$}lr@{$\: \pm \:$}lr@{$\: \pm \:$}lr@{$\: \pm \:$}l@{}}
\toprule										
Model	&	\multicolumn{2}{c}{A}	&	\multicolumn{2}{c}{B}	&	\multicolumn{2}{c}{C}	&	\multicolumn{2}{c}{D} \\
${}$	&	\multicolumn{2}{c}{Per cent}	&	\multicolumn{2}{c}{Per cent}	&	\multicolumn{2}{c}{Per cent}	&	\multicolumn{2}{c}{Per cent} \\
\midrule
B03$^{a}$	&	9.9	&	4.2	&	90.1	&	4.2	&	13.3	&	12.4	&	103.4	&	14.3	\\
D06$^{b}$	&	28.2	&	13.0	&	71.8	&	13.0	&	0.26	&	0.23	&	72.0 & 13.1	\\ [1ex]
B03	&	9.7	&	2.8	&	90.3	&	2.8	&	19.3	&	15.8	&	109.6	&	14.4	\\
D06	&	19.3	&	3.8	&	80.7	&	3.8	&	3.5	&	5.7	&	84.2	&	6.9	\\
P14	&	13.9	&	3.1	&	86.1	&	3.1	&	7.1	&	8.0	&	93.2	&	8.2	\\
\bottomrule											
\end{tabular}
\\ [1ex]
\raggedright
$^{a}$Data in this row is from Table 2 of \citetalias{Dunner06}.\\
$^{b}$Data in this row is from Table 1 of \citetalias{Dunner06}. \\
}
\end{table}

\begin{figure*}
\includegraphics[width=0.49\linewidth]{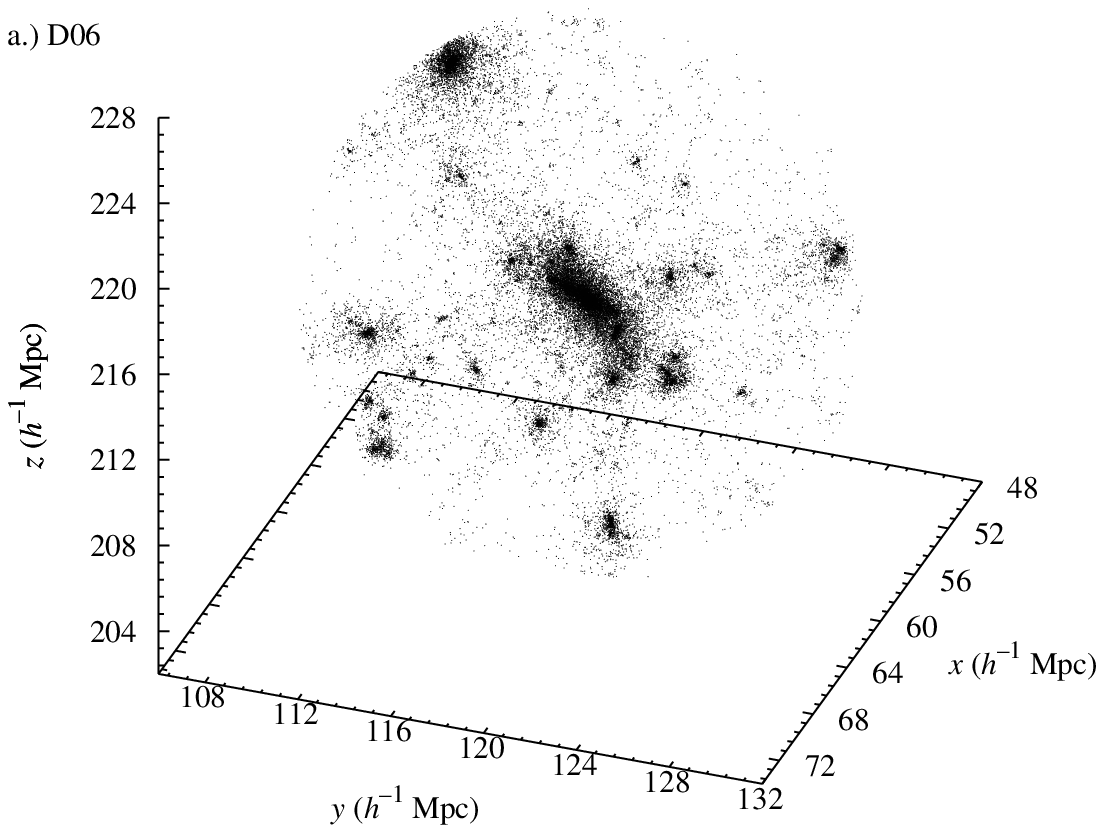} \hfill
\includegraphics[width=0.49\linewidth]{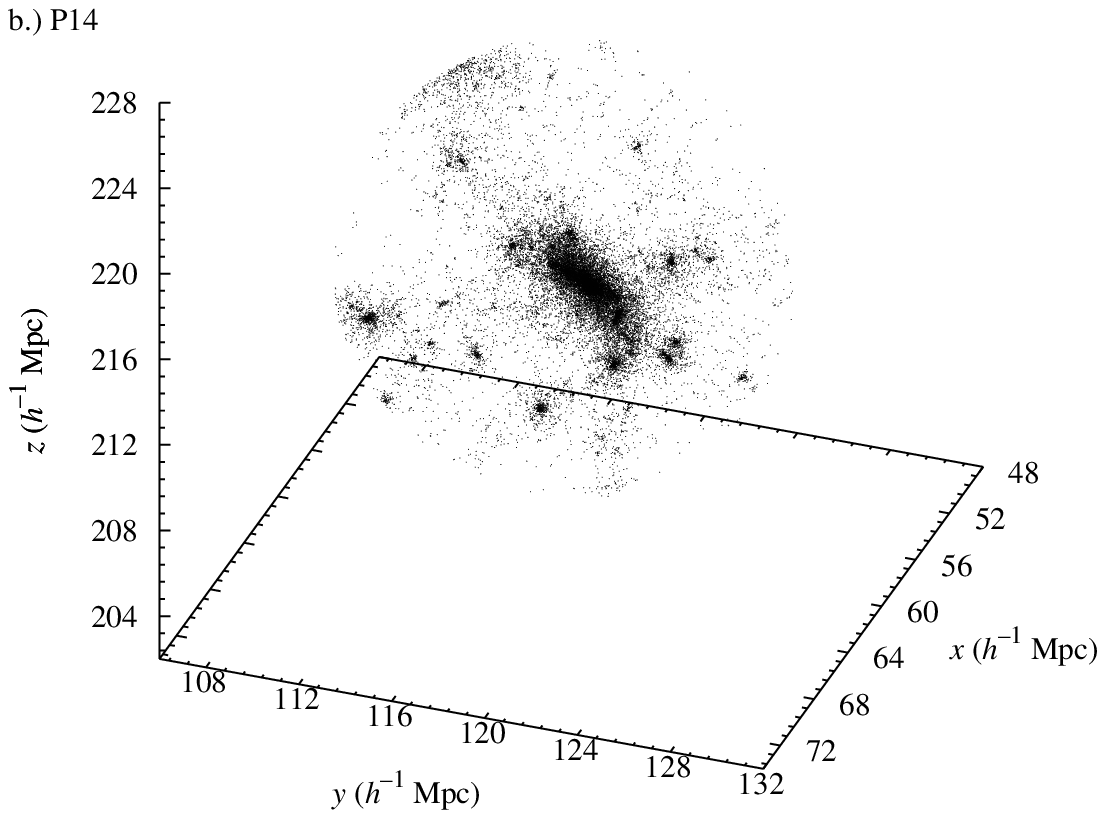} \hfill

\includegraphics[width=0.49\linewidth]{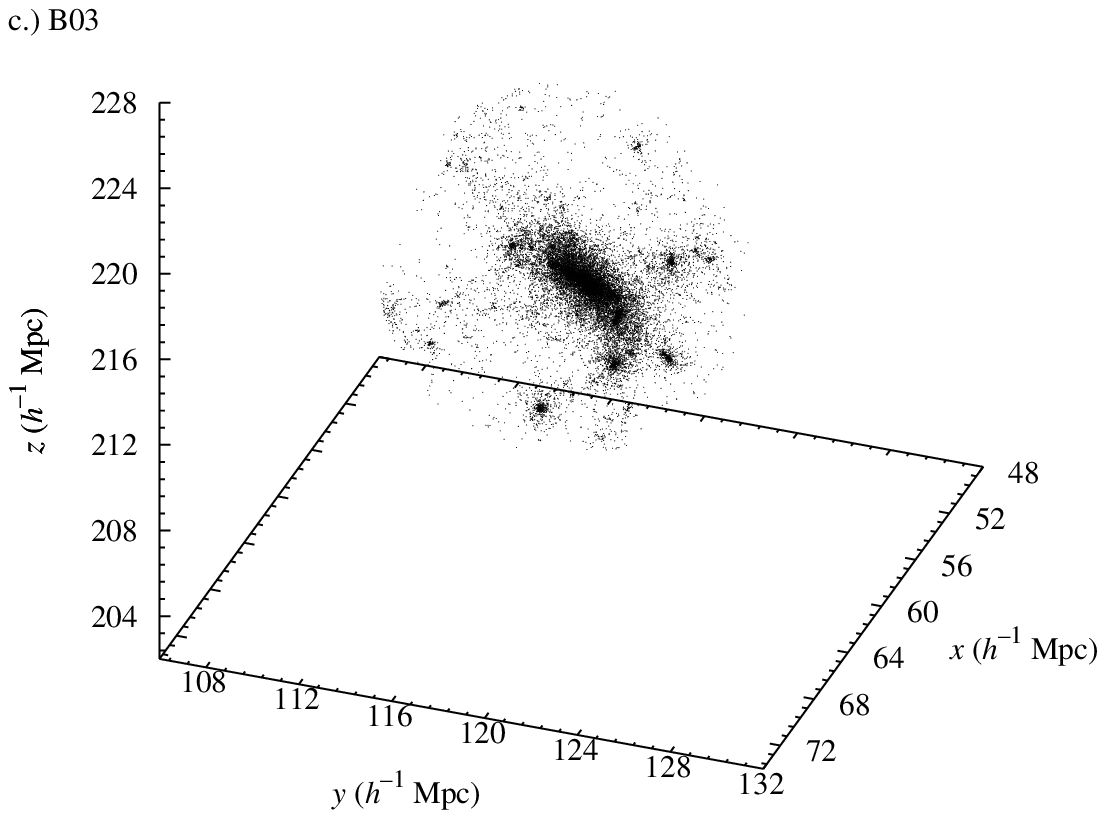} \hfill
\includegraphics[width=0.49\linewidth]{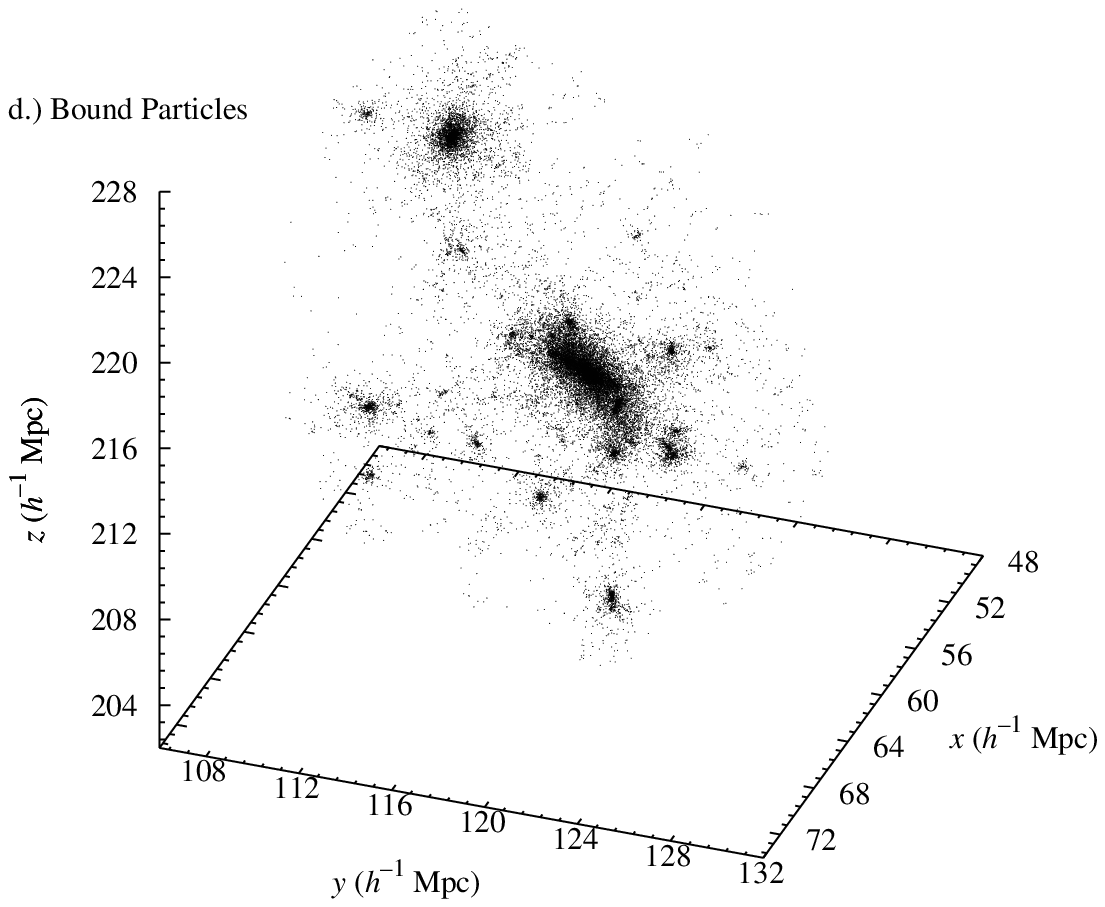}
\caption{Three-dimensional plots of the particles associated with Object 3 from Simulation 2. a.) All particles pulled from the present ($a=1$) snapshot based on the \citetalias{Dunner06} cutoff density. b.) All particles pulled from the present snapshot based on the P14 cutoff density. c.) All particles pulled from the present snapshot based on the \citetalias{Busha03} cutoff density. d.) All particles that are part of the structure in the future ($a=100$) snapshot in their present locations.}
\label{fig:Obj3}
\end{figure*}

Performing another visual examination of the objects gives the greatest insights into the accuracy of the models. Fig. \ref{fig:Obj3} shows Object 3 from Simulation 2. It can be seen that the \citetalias{Dunner06} model selects a cluster that is missed by the P14 model, and that this cluster is part of the structure in the future. The missed cluster's centre of mass lies \mbox{$\sim 1.5 \, h^{-1} \, \mathrm{Mpc}$} outside of the critical radius of the P14 model, making this the biggest failure of that model. It is worth noting that the central cluster is significantly elongated in the general direction of this missed cluster, giving it a very ellipsoidal shape. Since the density calculation used the mass enclosed by spherical shells, the non-spherical shape of the central cluster could explain the failure of the P14 model. It is also worth noting that the \citetalias{Busha03} model significantly underestimates the extent of the structure. For similar figures of all objects from this simulation, please see the online supplemental material.

\begin{figure}
\includegraphics[width=1.0\linewidth]{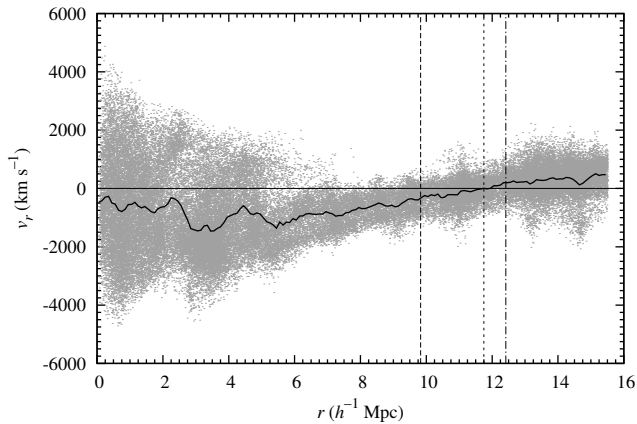}
\caption{Radial velocity versus distance from the central cluster for Object 1 in Simulation 2. The large dashed vertical line marks the turn around radius predicted by the P14 model. The dot-dashed vertical line marks the turn around radius predicted by the \citetalias{Dunner06} model. The gray dots are the radial velocities of individual particles, and the solid line is the average radial velocities in $100 \, h^{-1} \, \mathrm{kpc}$ bins. The small dashed vertical line marks where the average radial velocities transition from negative to positive, approximating the location of the true turn around radius.}
\label{fig:Vr}
\end{figure}

The velocity structure of objects in Simulation 2 was also examined to look for the location of the turn around radius. To do this, the centre of mass velocity for the central cluster was found and subtracted from the velocities of all the particles associated with that particular object. Then the velocities were broken into their radial and tangential components. Fig. \ref{fig:Vr} shows the radial velocities of particles around Object 1 in Simulation 2. The solid line represents the average radial velocities of particles in \mbox{$100 \, h^{-1} \mathrm{kpc}$} bins. From left to right, the vertical lines represent the turn around radii based on the P14 model, binned averages, and the \citetalias{Dunner06} model. The P14 model underestimates the turn around radius, while the \citetalias{Dunner06} model overestimates. The same trend is observed in the other objects as well, though for some, locating the turn around radius from the binned average is difficult due to clusters near that radius. On average, the P14 model underestimates the location of the turn around radius by \mbox{$\sim 16$} per cent, while the \citetalias{Dunner06} model overestimates its location by \mbox{$\sim 7$} per cent. This again suggests that the P14 model is slightly conservative, while the \citetalias{Dunner06} model is slightly optimistic.

\section{Applications to the CSC \& the Virgo Cluster}
It was recently shown by \citetalias{Pearson14} that five rich Abell clusters in the CSC are likely part of a gravitationally bound supercluster core. This provides an opportunity to test methods of applying these analytical models to a real structure in our Universe. The Fundamental Plane \citep{Djorgovski87} distance estimates published by \citetalias{Pearson14}, combined with the right ascensions and declinations allow for a reliable three-dimensional map of the CSC. In principle, it should be a straightforward task to calculate the critical radii of the various models by solving equation \eqref{eq:CritParam}, at the equality, for $r_{0}$ yielding
\begin{equation}
\label{eq:CriticalRadii}
r_{0} = \left(\dfrac{0.7^{2}M_{\mathrm{obj}}}{\xi \times 10^{12} \, \mathrm{M}_{\sun}}\right)^{1/3} \, h^{-1} \, \mathrm{Mpc}
\end{equation}
where $M_{\mathrm{obj}}$ is in units of \mbox{$h^{-1} \, \mathrm{M_{\sun}}$}. The only problem is that if only the virial mass is known, the critical radii will be underestimated, as additional mass outside of the virial radius will not be included. \cite{Rines2006} examined the caustics \citep{Diaferio99} around numerous clusters in the fourth data release of the Sloan Digital Sky Survey (SDSS) finding that, on average, the mass within turn around is \mbox{$2.19 \pm 0.18$} times the mass within the virial radius. \cite{Anderhalden2011}, by examining simulation snapshots before the present, located all particles that have collapsed at present but expanded beyond the virial radius after their first pericentric pass, finding that this would increase the mass of a halo by \mbox{$\sim 25$} per cent compared to the viral mass. Since their work did not include all the mass within turn around, this result is consistent with the work of \cite{Rines2006}. 

Using the objects in Simulation 1, $M_{200}$, the mass inside a radius enclosing 200 times the critical density, was used to calculate the critical radii via equation \eqref{eq:CriticalRadii} and compared with the value determined exactly using the enclosed density of particles in the simulation. On average the \citetalias{Busha03}, P14 and \citetalias{Dunner06} models using $M_{200}$ predicted radii that were \mbox{$77.9\pm 5.2$}, \mbox{$73.6 \pm 6.9$}, and \mbox{$73.9 \pm 7.6$} per cent of the values determined by the cutoff densities, respectively. Similar results are found using the objects in Simulation 2 with \mbox{$75.1 \pm 6.7$}, \mbox{$71.2 \pm 7.1$}, and \mbox{$72.0 \pm 7.2$} per cent, respectively.

The virial mass alone gives critical radii for A2065 of \mbox{$7.25_{-0.63}^{+0.17}$}, \mbox{$8.45_{-0.74}^{+0.19}$}, and \mbox{$9.89_{-0.86}^{+0.23} \, h^{-1} \, \mathrm{Mpc}$} for the \citetalias{Busha03}, P14, and \citetalias{Dunner06} models, respectively. The confidence intervals come directly from the uncertainty in the mass estimate of A2065 (\mbox{$2.33_{-0.56}^{+0.41} \times 10^{15} \, h^{-1} \, \mathrm{M}_{\sun}$}). If we assume that these are underestimates as indicated by the simulation data, the critical radii become \mbox{$9.3_{-1.1}^{+1.5}$}, \mbox{$11.4_{-1.8}^{+2.6}$}, and \mbox{$13.4_{-2.3}^{+3.5} \, h^{-1} \, \mathrm{Mpc}$} for the \citetalias{Busha03}, P14, and \citetalias{Dunner06} models, respectively. Here the confidence intervals come from the uncertainty of the degree of underestimation. It is worth noting that these results agree with the findings of \cite{Rines2006}. If the virial mass of A2065 is increased by a factor of \mbox{$2.19 \pm 0.18$}, the critical radii become \mbox{$9.41 \pm 0.26$}, \mbox{$11.0 \pm 0.3$} and \mbox{$12.8 \pm 0.3 \, h^{-1} \, \mathrm{Mpc}$} for the \citetalias{Busha03}, P14, and \citetalias{Dunner06} models, respectively. Looking at the separations that come from the FP distances given in Table \ref{tab:CSCSep}, we find that the model of P14 would include A2056, A2061, and A2089 inside the critical radius of A2065. However, A2061 and A2067 are only separated by $2.37 \, h^{-1} \, \mathrm{Mpc}$ and the mass of A2061 is $9.9_{-2.8}^{+2.2} \times  10^{14} \, h^{-1} \, \mathrm{M}_{\sun}$ giving it a critical radius of $\sim 9 \, h^{-1} \, \mathrm{Mpc}$ in the P14 model, taking into account the underestimation from the virial mass. This places A2067 well inside the critical radius of A2061 and as seen in section \ref{sec:GadgetComp}, everything that is bound to A2061 should also be included in the structure. This would imply that A2056, A2061, A2065, A2067, and A2089 should be a gravitationally bound structure, which is in complete agreement with the findings of \citetalias{Pearson14}. The model of \citetalias{Dunner06} would include all five of the above clusters inside the critical radius of A2065 alone.

\begin{table*}
\parbox{\linewidth}{
\caption{Relative separations of the clusters in the CSC from FP distances. Each column and row corresponds to a different cluster in the CSC. Matching one cluster's column to another's row gives their relative separation in units of $h^{-1} \, \mathrm{Mpc}$. Individual cluster distances have uncertainties of \mbox{$\sim 5$} per cent \citepalias{Pearson14}.}
\label{tab:CSCSep}
\centering
\begin{tabular}{@{}lcccccccc@{}}
\toprule
${}$	&	A2056	&	A2061	&	A2065	&	A2067	&	A2079	&	A2089	&	A2092	&	CL1529+29	\\
\midrule
A2056	&	--	&	9.08	&	4.63	&	10.48	&	22.60	&	11.18	&	16.57	&	30.83	\\
A2061	&	9.08	&	--	&	11.11	&	2.37	&	24.85	&	13.51	&	12.89	&	28.96	\\
A2065	&	4.63	&	11.11	&	--	&	12.41	&	25.09	&	8.83	&	17.90	&	27.55	\\
A2067	&	10.48	&	2.37	&	12.41	&	--	&	23.80	&	13.27	&	10.77	&	29.90	\\
A2079	&	22.60	&	24.85	&	25.09	&	23.80	&	--	&	22.60	&	17.51	&	50.51	\\
A2089	&	11.18	&	13.51	&	8.83	&	13.27	&	22.60	&	--	&	13.22	&	28.63	\\
A2092	&	16.57	&	12.89	&	17.90	&	10.77	&	17.51	&	13.22	&	--	&	36.36	\\
CL1529+29	&	30.83	&	28.96	&	27.55	&	29.90	&	50.51	&	28.63	&	36.36	&	--	\\
\bottomrule
\end{tabular}
}
\end{table*}

\begin{figure}
\includegraphics[width=1.0\linewidth]{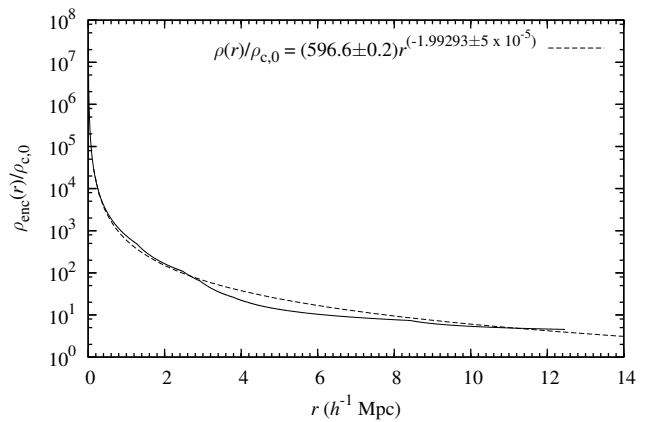}
\caption{Density versus radius as determined from an application of the caustics method to the CSC \citepalias{Pearson14}. The solid line are the data from the caustics calculation, and the dashed line is a power law fit. Since the ordinate values are normalized by the present value of the critical density, they are the shell density parameters.}
\label{fig:CSCDen}
\end{figure}

The above procedure has the advantage of not needing any details about the density profile of the structure. However, one has to assume that using $M_{200}$ from the simulation data is approximating the virial mass. Given the agreement with \cite{Rines2006}, this is likely not a bad assumption, but testing it further is still desirable. In applying the caustics method to a supercluster \citep{Reisenegger00}, the mass is determined as a function of radius. \citetalias{Pearson14} applied the caustics method to the CSC, giving mass as a function of radius which can be simply transformed into density as a function of radius. Fig. \ref{fig:CSCDen} shows the density normalized by the current value of the critical density (i.e. the density parameter) as a function of radius for the CSC from the caustics method. This was fit with a power law yielding
\begin{equation}
\Omega_{\mathrm{s}} = (596.6 \pm 0.2)r^{(-1.99293 \pm 5\times 10^{-5})}.
\end{equation}
This can then be solved for $r$ using the critical shell density parameters given in Table \ref{tab:ShellDens}. Doing this we obtain \mbox{$10.7$}, \mbox{$13.5$}, and \mbox{$16.1 \, h^{-1} \, \mathrm{Mpc}$} for the \citetalias{Busha03}, P14, and \citetalias{Dunner06} models, respectively. Taking the cluster separations from Table \ref{tab:CSCSep}, the P14 and \citetalias{Dunner06} models again predict that A2056, A2061, A2065, A2067, and A2089 are part of a gravitationally bound structure.

As a further test, the mass of the Virgo cluster within turn around was estimated using the P14 model. \cite{Karachentsev2014} examined the infall velocities of galaxies around the Virgo cluster. From their analysis they were able to determine that the turn around radius for Virgo is \mbox{$7.2 \pm 0.7 \, \mathrm{Mpc}$} (they assumed \mbox{$H_{0} = 72 \, \mathrm{km} \, \mathrm{s}^{-1} \, \mathrm{Mpc}^{-1}$}), and using the pure SCM (i.e. the \citetalias{Dunner06} model) they estimated the mass of the Virgo cluster to be \mbox{$(8.0 \pm 2.3) \times 10^{14} \, \mathrm{M}_{\sun}$}. They make note of the work of \cite{Rines2006}, but instead use the work of \cite{Anderhalden2011} to justify their findings as being in agreement with the virial mass of Virgo, \mbox{$(7.0 \pm 0.4) \times 10^{14} \, \mathrm{M}_{\sun}$}. Since \cite{Karachentsev2014} found the turn around radius, it would seem more appropriate to use the findings of \cite{Rines2006}, which would place the mass of Virgo within turn around at \mbox{$(1.5 \pm 0.2) \times 10^{15} \, \mathrm{M}_{\sun}$} based on the virial mass. Using the turn around density of the P14 model, a mass can be estimated for the Virgo cluster via
\begin{equation}
M_{\mathrm{ta}} = 7.17 \times 10^{12} \left(\dfrac{r_{\mathrm{ta}}}{\mathrm{Mpc}}\right)^{3} \, h^{2} \, \mathrm{M}_{\sun}.
\end{equation} 
With the turn around radius found by \cite {Karachentsev2014}, the mass estimate for Virgo would be \mbox{$(1.4 \pm 0.4) \times 10^{15} \, \mathrm{M}_{\sun}$}, in agreement with the work of \cite{Rines2006}.

\section{Discussion}
Identifying the extent of gravitationally bound structure could be very useful in placing constraints on models of our universe \citep{Sheth2011, Park2012}. Given the lack of spherical symmetry, effects of external attractors and tangential motions due to past gravitational interactions, setting the limits of bound structure is clearly not a straightforward task. Here, building on the work of \citetalias{Dunner06}, a simple modification of the SCM was made, yielding a new analytical model (P14) for the extent of bound structure in a universe containing cosmological constant dark energy. The P14 model was found to agree remarkably well with a model informed by the simulations of \citetalias{Pearson13}.

The two previous analytical models for the extent of bound structure, \citetalias{Busha03} and \citetalias{Dunner06}, along with the P14 model were tested against two \textsc{gadget2} simulations, which allowed for objects of a variety of sizes to be examined. By visually comparing the structures predicted as bound by the various models at present ($a=1$) to the structure as defined by the particles which remain in the far future ($a=100$), the \citetalias{Busha03} model was found to consistently miss portions of the structure. The \citetalias{Dunner06} model fairly consistently includes clusters which are not part of the structure in the far future. The P14 model performed well in many cases, though it does still seem to slightly underestimate the extent of bound structure. The fact that the P14 model performed well, and that it agrees so well with the \citetalias{Pearson13} informed model suggest that those simulations, while not very sophisticated, were still able to reliably locate bound structure as long as the peculiar motions along the line-of-sight are not significant, as they were in the case of the CSC \citepalias{Pearson14}. This would seem to suggest that by giving the clusters a Hubble expansion velocity at present, combined with open vacuum boundary conditions, the simulations of \citetalias{Pearson13} were in some ways modelling the effects of external attractors. 

In order to truly locate all structure that is bound, it is necessary to go beyond the simple spherical models. As seen here, after locating all the clusters that are predicted to be bound to the central cluster, if one then locates what is predicted to be bound to those clusters, it is possible to much more accurately determine what is part of the structure in the far future. This `non-spherical' approach is still fairly simple, and straightforward to apply. Future versions of \textsc{coldg}a\textsc{s} will do this automatically when locating potential gravitationally bound objects. Also, Object 3 from Simulation 2 (see Fig. \ref{fig:Obj3}) seems to suggest that when the central cluster itself is not very spherical, it may be necessary to alter the shape of the shells used in the calculation of the density to more closely resemble the central cluster. Another possible approach may be to build an analytical model for the limits of bound structure from an ellipsoidal collapse model (ECM). Recently, \cite{Rossi2011} used an ECM along with excursion set theory to study the shapes of dark matter haloes. The analytical model they derive was shown to more accurately model non-linear structure evolution than the Zeldovich approximation. Given the lack of agreement about how to define non-spherical gravitationally bound structures from simulation, they instead focus on comparing the axial ratios of structures in simulation to those predicted by their model. Strictly speaking, their work does not attempt to define the limits of bound structure, yet extending such an ECM could potentially provide a much more accurate analytical model which could inform the \textsc{coldg}a\textsc{s} software. It is hoped that further study of bound structures from simulation, as well as incorporating the lessons learned in this work, will lead to a version of \textsc{coldg}a\textsc{s} capable of reliably locating gravitationally bound structures in simulation.

Three different methods of applying these analytical models to real structures in our Universe have been shown. The first uses only the cluster virial masses, and makes some adjustment based on data from the simulations. In applying this method to the CSC, the P14 and \citetalias{Dunner06} models both agree with the results of \citetalias{Pearson14}. This method is definitely not ideal because of the assumption that \mbox{$M_{200}=M_{\mathrm{virial}}$}. However, given the agreement of the critical radii adjusted based on that assumption, and the critical radii from increasing the mass according to the results of \cite{Rines2006}, this should be a viable method for locating the extent of bound structure when very little data exists for the intercluster regions.

For structures that have been well studied, and have ample data available, knowledge gained from a caustics analysis may provide a better estimate of the extent of bound structure. The caustics analysis of the CSC \citepalias{Pearson14} was used to yield an estimate of the density profile of the structure. With this profile, it was a simple matter of locating the radii at which the various cutoff densities are reached. Using this method, the P14 and \citetalias{Dunner06} models again agree with the results of \citetalias{Pearson14}. 

Since the turn around densities of the P14 and \citetalias{Dunner06} models are known, if the turn around radius of a real structure can be reliably located, it is straightforward to then estimate the mass within turn around. \cite{Karachentsev2014} did just that with the pure SCM (\citetalias{Dunner06} model). Using the P14 model, the mass within the turn around radius of the Virgo cluster was found to be \mbox{$(1.4 \pm 0.4)\times 10^{15} \, \mathrm{M}_{\sun}$} which agrees very well with the findings of \cite{Rines2006}, that the total mass of a cluster inside the turn around radius is \mbox{$2.19 \pm 0.18$} times the virial mass.

Future work will examine these objects in greater detail, testing deviations from spherical shells for density calculations, performing an in depth examination of their velocity structure, as well as performing mock observations to gain a better sense of how these bound structures would appear in redshift survey data. Additionally, given that the true extent of bound structure seems to lie somewhere between that predicted by the P14 model and that predicted by the \citetalias{Dunner06} model, it should be possible to develop a model informed by the \textsc{gadget2} simulations performed here. In the end, this future work will seek to develop a method in the vein of \cite{Dunner07}, similar to the caustics method, in order to provide a mass estimate and dynamical description of structures.

\section*{Acknowledgements}
I would like to thank Merida Batiste whose FP analysis of the CSC was critical to parts of this work, and my Ph.D. adviser, David J. Batuski, who set me on a research path which led to this work. I would also like to acknowledge Daniel Eisenstein, whose comments on this portion of my Ph.D. dissertation helped to greatly strengthen this work.

\bibliographystyle{mn2emod}
\footnotesize
\bibliography{Pearson14}

\begin{thebibliography}{32}
\expandafter\ifx\csname natexlab\endcsname\relax\def\natexlab#1{#1}\fi

\bibitem[{{Anderhalden} \& {Diemand}(2011)}]{Anderhalden2011}
{Anderhalden} D., {Diemand} J., 2011, \mnras, 414, 3166

\bibitem[{{Araya-Melo} {et~al}\mbox{.}(2009){Araya-Melo}, {Reisenegger},
  {Meza}, {van de Weygaert}, {D{\"u}nner}, \& {Quintana}}]{Araya-Melo2009}
{Araya-Melo} P.~A., {Reisenegger} A., {Meza} A., {van de Weygaert} R.,
  {D{\"u}nner} R., {Quintana} H., 2009, \mnras, 399, 97

\bibitem[{{Bardelli} {et~al}\mbox{.}(1993){Bardelli}, {Scaramella},
  {Vettolani}, {Zamorani}, {Zucca}, {Collins}, \&
  {MacGillivray}}]{Bardelli1993}
{Bardelli} S., {Scaramella} R., {Vettolani} G., {Zamorani} G., {Zucca} E.,
  {Collins} C.~A., {MacGillivray} H.~T., 1993, The Messenger, 71, 34

\bibitem[{{Batiste} \& {Batuski}(2013)}]{Batiste13}
{Batiste} M., {Batuski} D.~J., 2013, \mnras, 436, 3331

\bibitem[{{Beers}, {Flynn} \& {Gebhardt}(1990){Beers}, {Flynn}, \&
  {Gebhardt}}]{Beers90}
{Beers} T.~C., {Flynn} K., {Gebhardt} K., 1990, \aj, 100, 32

\bibitem[{{Busha} {et~al}\mbox{.}(2003){Busha}, {Adams}, {Wechsler}, \&
  {Evrard}}]{Busha03}
{Busha} M.~T., {Adams} F.~C., {Wechsler} R.~H., {Evrard} A.~E., 2003, \apj,
  596, 713

\bibitem[{{Colpi}, {Mayer} \& {Governato}(1999){Colpi}, {Mayer}, \&
  {Governato}}]{Colpi1999}
{Colpi} M., {Mayer} L., {Governato} F., 1999, \apj, 525, 720

\bibitem[{{Diaferio}(1999)}]{Diaferio99}
{Diaferio} A., 1999, \mnras, 309, 610

\bibitem[{{Djorgovski} \& {Davis}(1987)}]{Djorgovski87}
{Djorgovski} S., {Davis} M., 1987, \apj, 313, 59

\bibitem[{{D{\"u}nner} {et~al}\mbox{.}(2006){D{\"u}nner}, {Araya}, {Meza}, \&
  {Reisenegger}}]{Dunner06}
{D{\"u}nner} R., {Araya} P.~A., {Meza} A., {Reisenegger} A., 2006, \mnras, 366,
  803

\bibitem[{{D{\"u}nner} {et~al}\mbox{.}(2007){D{\"u}nner}, {Reisenegger},
  {Meza}, {Araya}, \& {Quintana}}]{Dunner07}
{D{\"u}nner} R., {Reisenegger} A., {Meza} A., {Araya} P.~A., {Quintana} H.,
  2007, \mnras, 376, 1577

\bibitem[{{Hoffman} {et~al}\mbox{.}(2007){Hoffman}, {Lahav}, {Yepes}, \&
  {Dover}}]{Hoffman2007}
{Hoffman} Y., {Lahav} O., {Yepes} G., {Dover} Y., 2007, \jcap, 10, 16

\bibitem[{{Karachentsev} {et~al}\mbox{.}(2014){Karachentsev}, {Tully}, {Wu},
  {Shaya}, \& {Dolphin}}]{Karachentsev2014}
{Karachentsev} I.~D., {Tully} R.~B., {Wu} P.-F., {Shaya} E.~J., {Dolphin}
  A.~E., 2014, ApJ, 782, 4

\bibitem[{{Lahav} {et~al}\mbox{.}(1991){Lahav}, {Lilje}, {Primack}, \&
  {Rees}}]{Lahav1991}
{Lahav} O., {Lilje} P.~B., {Primack} J.~R., {Rees} M.~J., 1991, \mnras, 251,
  128

\bibitem[{{Luparello} {et~al}\mbox{.}(2011){Luparello}, {Lares}, {Lambas}, \&
  {Padilla}}]{Luparello2011}
{Luparello} H., {Lares} M., {Lambas} D.~G., {Padilla} N., 2011, \mnras, 415,
  964

\bibitem[{{Mu{\~n}oz} \& {Loeb}(2008)}]{Munoz2008}
{Mu{\~n}oz} J.~A., {Loeb} A., 2008, \mnras, 391, 1341

\bibitem[{{Nagamine} \& {Loeb}(2003)}]{Nagamine2003}
{Nagamine} K., {Loeb} A., 2003, \na, 8, 439

\bibitem[{{Park} {et~al}\mbox{.}(2012){Park}, {Choi}, {Kim}, {Gott}, {Kim}, \&
  {Kim}}]{Park2012}
{Park} C., {Choi} Y.-Y., {Kim} J., {Gott}, III J.~R., {Kim} S.~S., {Kim} K.-S.,
  2012, \apjl, 759, L7

\bibitem[{{Pearson}, {Batiste} \& {Batuski}(2014){Pearson}, {Batiste}, \&
  {Batuski}}]{Pearson14}
{Pearson} D.~W., {Batiste} M., {Batuski} D.~J., 2014, \mnras, 441, 1601

\bibitem[{{Pearson} \& {Batuski}(2013)}]{Pearson13}
{Pearson} D.~W., {Batuski} D.~J., 2013, \mnras, 436, 796

\bibitem[{{Peebles}(1980)}]{Peebles1980}
{Peebles} P.~J.~E., 1980, {The large-scale structure of the universe}.
  Princeton University Press, Princeton, NJ

\bibitem[{{Percival}(2005)}]{Percival2005}
{Percival} W.~J., 2005, \aap, 443, 819

\bibitem[{{Planck Collaboration} {et~al}\mbox{.}(2014){Planck Collaboration},
  {Ade}, {Aghanim}, {Armitage-Caplan}, {Arnaud}, {Ashdown}, {Atrio-Barandela},
  {Aumont}, {Baccigalupi}, {Banday}, \& et~al.}]{Ade2014}
{Planck Collaboration} {et~al.}, 2014, \aap, 571, A16

\bibitem[{{Proust} {et~al}\mbox{.}(2006){Proust}, {Quintana}, {Carrasco},
  {Reisenegger}, {Slezak}, {Muriel}, {D{\"u}nner}, {Sodr{\'e}}, {Drinkwater},
  {Parker}, \& {Ragone}}]{Proust06}
{Proust} D. {et~al.}, 2006, \aap, 447, 133

\bibitem[{{Reisenegger} {et~al}\mbox{.}(2000){Reisenegger}, {Quintana},
  {Carrasco}, \& {Maze}}]{Reisenegger00}
{Reisenegger} A., {Quintana} H., {Carrasco} E.~R., {Maze} J., 2000, \aj, 120,
  523

\bibitem[{{Rines} \& {Diaferio}(2006)}]{Rines2006}
{Rines} K., {Diaferio} A., 2006, \aj, 132, 1275

\bibitem[{{Rossi}, {Sheth} \& {Tormen}(2011){Rossi}, {Sheth}, \&
  {Tormen}}]{Rossi2011}
{Rossi} G., {Sheth} R.~K., {Tormen} G., 2011, \mnras, 416, 248

\bibitem[{{Sheth} \& {Diaferio}(2011)}]{Sheth2011}
{Sheth} R.~K., {Diaferio} A., 2011, \mnras, 417, 2938

\bibitem[{{Small} {et~al}\mbox{.}(1998){Small}, {Ma}, {Sargent}, \&
  {Hamilton}}]{Small98}
{Small} T.~A., {Ma} C.-P., {Sargent} W.~L.~W., {Hamilton} D., 1998, \apj, 492,
  45

\bibitem[{{Springel}(2005)}]{Springel05}
{Springel} V., 2005, \mnras, 364, 1105

\bibitem[{{Wang} \& {Steinhardt}(1998)}]{Wang1998}
{Wang} L., {Steinhardt} P.~J., 1998, \apj, 508, 483

\bibitem[{{White}(1978)}]{White1978}
{White} S.~D.~M., 1978, \mnras, 184, 185

\end{thebibliography}

\label{lastpage}

\end{document}